\documentclass[lettersize,journal]{IEEEtran}
\usepackage{amsmath,amsfonts}
\usepackage{algorithmic}
\usepackage{array}
\usepackage{stackengine}
\usepackage[caption=false,font=normalsize,labelfont=sf,textfont=sf]{subfig}
\usepackage{textcomp}
\usepackage{stfloats}
\usepackage{mathtools}
\usepackage{url}
\usepackage{verbatim}
\usepackage{graphicx}
\usepackage{amssymb}
\usepackage[ruled]{algorithm2e}
\usepackage{lipsum}
\usepackage{multirow, booktabs}
\usepackage{array}
\usepackage{float}
\def\BibTeX{{\rm B\kern-.05em{\sc i\kern-.025em b}\kern-.08em
		T\kern-.1667em\lower.7ex\hbox{E}\kern-.125emX}}
\usepackage{balance}
\usepackage{blindtext, xcolor}
\usepackage{nomencl}
\usepackage{wasysym}

\newtheorem{theorem}{Theorem}

\newtheorem{proposition}{Proposition}
\newtheorem{lemma}{Lemma}
\begin{document}
	\title{Learning Robust Model Predictive Control for Voltage Control of Islanded Microgrid}
\author{Sahand~Kiani,          Hamed~Kebriaei$^*$,~\IEEEmembership{Senior Member,~IEEE}, Mohsen~Hamzeh, and  Ali~Salmanpour 
\thanks{ $^*$Corresponding Author (Hamed Kebriaei)}
\thanks{S. Kiani, H. Kebriaei, M. Hamzeh, A. Salmanpour are with the School of Electrical and Computer Engineering, College of Engineering, University of Tehran, Tehran, Iran. (emails: sahand.kiani1374@gmail.com, kebriaei@ut.ac.ir, mohsenhamzeh@ut.ac.ir, a.salmanpour9@gmail.com)}}
	
	\maketitle
	
	\begin{abstract}
		This paper proposes a novel control design for voltage tracking of an islanded AC microgrid in the presence of {nonlinear} loads and parametric uncertainties at the primary level of control. The proposed method is based on the Tube-Based Robust Model Predictive Control (RMPC), an online optimization-based method which can handle the constraints and uncertainties as well. The challenge with this method is the conservativeness imposed by designing the tube based on the worst-case scenario of the uncertainties. This weakness is amended in this paper by employing a combination of a learning-based Gaussian Process (GP) regression and RMPC. The advantage of using GP is that both the mean and variance of the loads are predicted at each iteration based on the real data, and the resulted values of mean and the bound of confidence are utilized to design the tube in RMPC. The theoretical results are also provided to prove the recursive feasibility and stability of the proposed learning based RMPC. 
        Finally, the simulation results are carried out on both single and multiple DG (Distributed Generation) units. 

		\textit{Note for Practitioners}-- In this paper, we present a new way to control the voltage in an islanded microgrid to improve Power Quality (PQ). The method we propose is based on an online optimization technique called Tube-Based Robust Model Predictive Control. It can handle uncertainties and disturbances that occur when the microgrid operates independently, ensuring the voltage remains stable. However, there's a challenge with this method. It tends to be too cautious because it assumes the worst-case scenario for uncertainties. To make the control more efficient, we improve it by combining a learning-based technique called Gaussian Process regression with RMPC. The advantage of using GP is that it predicts the uncertainty of the electrical devices based on real data. We use these predictions to design the control in RMPC more accurately. We also provide theoretical results to show that our new learning-based control is reliable and stable. We tested our approach through computer simulations on different scenarios with one or multiple power sources in the microgrid. The results show the effectiveness of our control design in regulating the voltage even with uncertain and nonlinear loads. Overall, this paper suggests a practical and reliable way to control the voltage in an independent microgrid using a combination of online optimization and learning techniques.
		
	\end{abstract}
 
	\begin{IEEEkeywords}
		Gaussian Process Regression, Islanded Microgrid, Robust Model Predictive Control, Voltage Control. 
	\end{IEEEkeywords}

	\section{Introduction}
	
    \IEEEPARstart{S}{}\hspace{7pt}
    everal reasons make the microgrids work in an islanded mode, such as faults in the main grid, high prices of grid's power, and supplying remote areas \cite{dehkordi2016fully}. Improving local reliability, power quality, providing lower investment costs, and reducing emissions are microgrids' main features \cite{singh2014microgrid}. Voltage regulation is a vital requirement to maintain the power delivery stable and consistent in an islanded microgrid. Proper voltage regulation ensures reliable and efficient operation of the microgrid, especially in remote and isolated areas where grid connections are not available \cite{bidram2014distributed}.

    Traditional voltage regulation methods, such as PI control, fuzzy logic control, sliding mode control, and adaptive control have been widely used in microgrids \cite{shen2023cascade}\cite{thomas2021fuzzy}. PI control is a simple and reliable method that provides good stability, but it has limited performance when dealing with nonlinear and uncertain systems \cite{merritt2017new}. Fuzzy logic control is a more flexible method that can handle nonlinear and uncertain systems but requires more complex modeling and control design \cite{huang2018model}. In recent years, researchers have focused on developing advanced control strategies, such as MPC, to improve the performance and stability of islanded microgrids \cite{9107341}\cite{7801066}. MPC is an optimization-based control method that can handle both linear and nonlinear systems with constraints, making it more suitable for complex and uncertain systems. It has been shown that in terms of steady-state performance, MPC outperforms sliding mode control \cite{doostdar2017comparing}.   
    Explicit prediction of future plant behavior and the computational efficiency of MPC using current systems measurements made this online method favorable in microgrids \cite{8704275}\cite{sockeel2020virtual}. In \cite{zhang2019model}, the authors conducted simulations and experimental tests to compare the performance of the proposed MPC method with traditional control methods such as PI control and fuzzy logic control. The results showed that the MPC-based control method outperformed traditional methods in terms of voltage regulation performance, especially under conditions of varying loads and disturbances.
     
    To face with the uncertainties and disturbances existing in the model, Tube-Based RMPC is proposed \cite{alessio2015tube}. 
    The feedback law tights the real state trajectories inside a bound centered around the nominal system trajectories. However, this method can be conservative since the control law is obtained based on the worst-case scenario of uncertainties \cite{oshnoei2023robust}.
    
    To mitigate conservativeness in Tube-Based RMPC, various regression methods have been proposed to improve the accuracy of the tube approximation \cite{paulson2011tube}. One such method is Gaussian Process Regression (GPR). GPR is a supervised learning-based method that has shown promise in estimating and compensating for disturbances in MPC while reducing conservativeness and improving computational efficiency \cite{Zhang2020}. It has also been used in power systems studies  for distribution system voltage control \cite{li2020mpc}. GPR estimates a bound of confidence of prediction, in addition to the mean value, which is directly used in robust controller design \cite{hewing2020learning}. 
    By using GPR to estimate disturbances in real time, the performance of RMPC is enhanced by reducing the need for conservative control strategies.

     In this paper, we propose the Learning RMPC method that utilizes GPR to estimate the current of the load as the disturbance, while also considering the upper bound of parametric uncertainties in the robust controller design. {To the best of the authors' knowledge, neither Tube-Based RMPC nor Learning RMPC has been used to control the voltage of an islanded microgrid at the primary level.} The proposed method is initially developed for a single-DG islanded microgrid, where nonlinear loads and parametric uncertainties can significantly affect the system's performance and power quality. Nevertheless, we also demonstrate its effectiveness for multiple DGs as well. Since the control system of an islanded microgrid is hierarchical with different layers,
    in the case of multiple DG units, power sharing control loop is added. {Moreover, the recursive feasibility and stability of the proposed method have been verified analytically.}
The main contributions of this paper are:
	
	\begin{itemize}
		\item Using Tube-Based RMPC to enhance the performance of voltage regulation of an islanded microgrid in the presence of parametric uncertainties and disturbances (including harmonic loads) to meet PQ standards
		\item Using Gaussian Process as a learning method to calculate the tubes online with less conservativeness and estimate the loads to be used in shaping the control law
		\item Proposing recursive feasibility and stability analysis for the system, in which the effect of GP is considered
        \item {Demonstrating the effectiveness of the proposed control method at the primary level while multiple DGs exist in an islanded microgrid} 
	\end{itemize}
	
	
	\textbf{Notation}: 
 The Euclidean norm of $x$ vector is denoted by $\|x\|$. Respectively, The term $x^TQx$ denotes $\|x\|^2_Q$. The Pontryagin difference of sets $A \subseteq \mathbb{R}^n$ and $B \subseteq \mathbb{R}^n$ is denoted by $A \ominus B = \{a|a+b \in A, \forall b \in B\}$, and the Minkowski sum is $A \oplus B = \{a+b|a \in A, b \in B\}$. We define the addition of set and vector as $x \oplus B = \{x+b|x \in \mathbb{R}^n, b \in B\}$. The set multiplication is defined as if $M \in R^{m \times n}$ then $MA := \{Ma|a \in A\}$. $I_n$ is used to show the identity matrix of size n. The pseudo-inverse of a $m \times n$ matrix $X$ when $n \le m$ is $A^{\dagger}=(A^TA)^{-1}A^T$. The set $\psi \subseteq \mathbb{R}^n$ is positive invariant if and only if no solution starting inside $\psi$ can leave $\psi$. i.e. $\forall x(0) \in \psi,\,\, \varphi(k,x(0)) \in \psi\,\, \forall k > 0$, for dynamical system ${x(k+1)} = f(x(k))$ with $\varphi(k,x(0))$ as its solution.
	
	
	\section{Single-DG In An Islanded Microgrid}\label{section:m}
	
    A single-DG in an islanded microgrid is considered. Overhauls, faults, high energy prices, and supplying remote areas are among the reasons to switch a microgrid from the grid-connected mode to the islanded mode. Various disturbances might arise when the microgrid is in islanded mode, such as current of loads. The parametric uncertainties in the inverter's LC filter are another source of disturbance. This paper takes into account the effect of parametric uncertainties and loads as disturbances to the microgrid and proposes an appropriate control method to compensate for such effects. The microgrid is defined by three main components in islanded mode: DG, converter, and loads.

	\begin{figure}[H]
		\begin{center}
			\includegraphics[width=8.75cm]{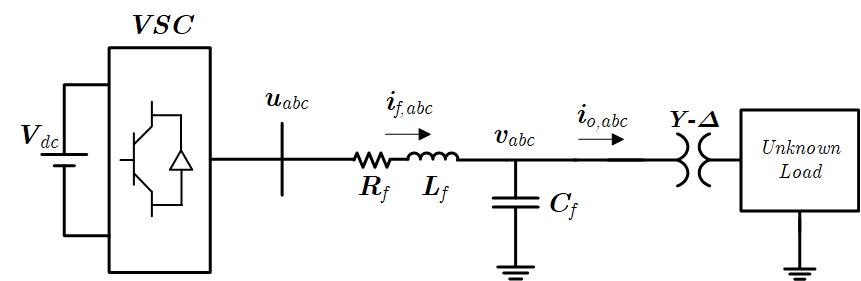}
			\caption{Single-DG unit connected to an unknown load}
			\label{fig:DGunit}
		\end{center}
	\end{figure}
 
	As shown in Fig. \ref{fig:DGunit}, by neglecting the nonlinear dynamics of the DC-Bus and using Kirchhoff’s voltage and current laws, the dynamical model can be derived as follows
	
	\begin{equation}
	\begin{cases}
	i_{f,abc} = i_{o,abc} + C_{f}\frac{dv_{abc}}{dt}\\
	u_{abc} = L_f\frac{di_{f,abc}}{dt}+R_fi_{f,abc}+v_{abc}\label{eq:1}
	\end{cases}
	\end{equation}
	where $i_{f,abc}$ and $i_{o,abc}$ stand for three-phase inductor current and inverter output current. respectively, $u_{abc}$ and $v_{abc}$ show the inverter output voltage before LC filter and output terminal voltage. Moreover, $R_f$, $L_f$, and $C_f$ are the parameters of the filter. To exploit the advantages of the d-q frame for smoother analysis, Park transformation is used as follows
	
	\begin{equation}
	\begin{bmatrix}
	\dot{V_d} \\
	\dot{V_q} \\
	\dot{I}_{fd} \\
	\dot{I}_{fq}
	\end{bmatrix}
	= 
	\begin{bmatrix}
	\omega_0V_q+\frac{1}{C_f}I_{fd}-\frac{1}{C_f}I_{od} \\
	-\omega_0V_d+\frac{1}{C_f}I_{fq}-\frac{1}{C_f}I_{oq} \\
	-\frac{1}{L_f}V_d-\frac{R_f}{L_f}I_{fd}+\omega_0I_{fq}+\frac{1}{L_f}V_{od}\\
	-\frac{1}{L_f}V_q-\omega_0I_{fd}-\frac{R_f}{L_f}I_{fq}+\frac{1}{L_f}V_{oq}
	\end{bmatrix}\label{eq:4}
	\end{equation}
	
	Equation \eqref{eq:4} forms the small-signal dynamical model of a single-DG unit in d-q frame. To make the model compatible with MPC, the model is discretized so that the discrete state-space model can be expressed as follows 
	
	\begin{equation}
	\begin{aligned}
	&x(k+1) = Ax(k)+Bu(k)+\underbrace{Eu_d(k)}_{w_1(k)}+\underbrace{\Delta Ax(k)+\Delta Bu(k)}_{w_2(x,u,k)} \\[-15pt]
	&y = Cx(k),
	\end{aligned}\label{eq:ModelDisc}
	\end{equation}
	where $x = [V_d\; V_q\; I_{fd}\; I_{fq}]^T\in X \subseteq R^4$ denotes the state vector and $u = [V_{od}\; V_{oq}]^T \in U \subseteq R^2, u_d = [I_{od}\; I_{oq}]$ are control and disturbance input, respectively, and $y = [V_d\; V_q] \in R^2$ is the output vector. The vectors $w_1(k)$ and $w_2(x,u,k)$ represent the effect of the current of the loads and parametric uncertainties in the 3-phase lines, respectively. During the operation of an inverter, the capacitance, the inductance, and the resistance of the LC filter vary as time passes. These changes are bounded, unpredictable, and adversely affect voltage tracking. This effect is captured by the uncertainty matrices $\Delta A$ and $\Delta B$, obtained by substituting $C_f+\Delta C_f$ for $C_f$, $L_f+\Delta L_f$ for $L_f$, and $R_f+\Delta R_f$ for $R_f$ in \eqref{eq:4}, respectively.  The matrices will be defined as
	
	\begin{align}
	A& =
	\begin{bmatrix}
	0 & \omega_0 & \frac{1}{C_f} & 0 \\
	-\omega_0 & 0 & 0 & \frac{1}{C_f} \\
	-\frac{1}{L_f} & 0 & -\frac{R_f}{L_f} & \omega_0 \\
	0 & -\frac{1}{L_f} & -\omega_0 & -\frac{R_f}{L_f}
	\end{bmatrix},
	B =
	\begin{bmatrix}
	0 & 0 \\
	0 & 0 \\
	\frac{1}{L_f} & 0\\
	0 & \frac{1}{L_f}
	\end{bmatrix}\nonumber
	\\E&=
	\begin{bmatrix}
	-\frac{1}{C_f} & 0 \\
	0 & -\frac{1}{C_f} \\
	0 & 0\\
	0 & 0
	\end{bmatrix},\nonumber
	C= 
	\begin{bmatrix}
	1 & 0 & 0 & 0 \\
	0 & 1 & 0 & 0 \\
	\end{bmatrix}\nonumber
	\end{align}
	
	This paper follows a novel approach to facing uncertainties and disturbances stated in \eqref{eq:ModelDisc}. $w_1(k)$, known as the current of the load, is the measured disturbance. GP estimator is utilized to learn the mean and the variance of $w_1(k)$. On the other hand, the upper bound for the parametric uncertainty $w_2(x,u,k)$ is predetermined and known, i.e. $\|w_2(x,u,k)\|_\infty \le L_{2}$. It should be noted that the pattern of parametric uncertainties related to the LC filter depends on the materials used in the inductor and capacitance, electric flux, and temperature. As a result, in order to estimate these values, the data should be gathered in laboratory circumstances or from the designer. Therefore, we only use a worst-case bound for parametric uncertainty $w_2(x,u,k)$. The estimated mean and variance of $w_1(k)$ and the bound of $w_2(x,u,k)$ will be used to design Tube-Based RMPC. In what follows, we introduce the GP estimator of load $w_1(k)$ and Tube-Based RMPC as two main components of the proposed scheme.
\vspace{5pt}
 
	\section{Preliminary Results}\label{section:PR}
	
	\subsection{Load's current prediction using Gaussian Process}
	
	Loads are considered the most regular disturbances in islanded microgrids. The current of the loads can be looked as a time series and estimated at each time instant. Here, GPR is employed as one of the most useful Bayesian nonparametric models. By using GPR, not only an estimate from the load current is attained, but also we can estimate a confidence interval which will be used as a bound in designing Tube-Based RMPC. Since we estimate this bound using real data, it would be less conservative compared to the worst-case bound. 
	
	Consider the standard linear regression model with Gaussian noise
	\begin{equation}
	\tilde{w}_1(k_i) = w_1(k_i) + \varepsilon (k_{i})
	\end{equation}
    where $\widetilde{\omega}_{1} (k_{i})$ denotes measured value, $ \varepsilon (k_{i}) $ represents an additive noise. $\omega_{1} (k_{i})$  is the disturbance of the DG at time instant $k_{i}$. The noise $\varepsilon\sim N(0,\delta^2_{\varepsilon})$ has the property of independent and identically distributed (i.i.d) Gaussian distribution with zero mean and variance $\delta^2_{\varepsilon}$. We introduce $\widetilde{\textbf{w}}_{1}(\tau) = [\widetilde{\omega}_{1} (k_{1}),\widetilde{\omega}_{1} (k_{2}),...,\widetilde{\omega}_{1} (k_{n})]$ as measured value vector, $\tau = [k_{1},k_{2},...,k_{n}]$ as the sample time vector and ${\textbf{w}}_{1}(\tau) = [{\omega}_{1} (k_{1}),{\omega}_{1} (k_{2}),...,{\omega}_{1} (k_{n})]$ as load disturbance vector.
     The aim is to assign a distribution over ${\textbf{w}}_{1}(\tau)$ given the measured values of $\widetilde{\textbf{w}}_{1}(\tau)$  using GP \cite{books/lib/RasmussenW06}. The distribution function of the measured vector is considered to be drawn from a multivariate Gaussian distribution as follows \cite{roberts2013gaussian}

	\begin{equation}
	p(\tilde{\textbf{w}}_1(\tau)) = \mathcal{N}(\mu(\tau),{C}(\tau,\tau))
	\end{equation}
	where $\mu(\tau) = [m(k_1),\dots,m(k_n)]$ is denoted as the vector of mean functions $m(\tau) = \mathbb{E}[\tilde{w}(\tau)]$. ${C}(\tau,\tau)$ defined as follows
	
	\begin{equation}
	{C}(\tau,\tau) = 
	\begin{pmatrix}
	c(k_1,k_1) & c(k_1,k_2) & \dots & c(k_1,k_n)\\
	c(k_2,k_1) & c(k_2,k_2) & \dots & c(k_2,k_n)\\
	\vdots & \vdots & \vdots & \vdots	\\
	c(k_n,k_1) & c(k_n,k_2) & \dots & c(k_n,k_n)
	\end{pmatrix}
	\end{equation}
	is known as the covariance matrix whose elements are covariance kernel functions between two training inputs. The kernel function is the degree of freedom that plays an essential role in the estimation. As it has been discussed in \cite{roberts2013gaussian}, choosing the kernel function depends on the knowledge of data. In this paper, the Radial Basis Function (RBF), also known as \textit{Squared Exponential} (\textit{SE}) kernel, is used as the kernel function $c(k_i,k_j)=h^2exp[-(\frac{k_i-k_j}{\lambda})^2]$ in the covariance matrix of GP. The \textit{SE} kernel has two hyperparameters. The length scale $\lambda$ determines the length of the wiggle in the function, and the output-scale $h$ determines the average distance of the function away from the mean.

	In order to generate a prediction using GPR, we need to compute the posterior distribution function of $\bf{w_1}$ and then predict unseen data by calculating the predictive posterior \cite{swastanto2016gaussian}. The posterior function consists of two components; likelihood function and prior distribution. The likelihood function is defined as follows
	\begin{equation} 
	p(\tilde{\textbf{w}}_1(\tau)|\textbf{w}_1(\tau)) \sim \mathcal{N}(\tilde{\textbf{w}}_1(\tau)|\textbf{w}_1(\tau),\sigma^2_NI)\label{eq:likelihoodFun}
	\end{equation} 
	where the mean of the function is centered on an arbitrary $\textbf{w}_1$. The second component is the prior defined as follows
	\begin{equation} \label{eq9}
	p(\textbf{w}_1(\tau)|\tau) \sim \mathcal{N}(\textbf{w}_1(\tau)|\textbf{0},C(\tau,\tau))
	\end{equation}
	
	It is assumed the mean of the prior is always a zero vector. Nevertheless, as we see in what follows, GP is able to model a general mean function from the kernel-based covariance function. From \eqref{eq:likelihoodFun} and \eqref{eq9}, the posterior over function is achieved by the likelihood function times the prior over function
	\begin{equation}
	p(\textbf{w}_1(\tau)|\tau,\tilde{\textbf{w}}_1(\tau)) \propto p(\tilde{\textbf{w}}_1(\tau)|\textbf{w}_1(\tau))p(\textbf{w}_1(\tau)|\tau)
	\end{equation}
	
	Both the likelihood and the prior are Gaussian. As a result, posterior over function will have a Gaussian distribution as follows
	
	\begin{equation}
	\begin{aligned}
	&p(\textbf{w}_1(\tau)|\tau,\tilde{\textbf{w}}_1(\tau)) \sim \mathcal{N}(\textbf{w}_1(\tau)|\bar{\mu},\bar{\Sigma})\\
	&\bar{\mu} = {C}(\tau,\tau)[{C}(\tau,\tau)+\sigma^2_NI]^{-1}\tilde{\textbf{w}}_1(\tau)\\
	&\bar{\Sigma} = {C}(\tau,\tau)[{C}(\tau,\tau)+\sigma^2_NI]^{-1}\sigma^2_NI
	\end{aligned}\label{eq:posteriorfun}
	\end{equation}
	
	Once the posterior over function is obtained, The predictive posterior can be computed as follows, which is used to predict the mean and variance of unseen data.
	
	\begin{equation}
	\begin{aligned}
	p&(\tilde{\textbf{w}}_1^{\ast}(\tau)|\tau_{\ast},\tau,\tilde{\textbf{w}}_1(\tau)) = \\
	&\int p(\tilde{\textbf{w}}_1^{\ast}(\tau)|\tau_{\ast},\textbf{w}_1(\tau),\tau)p(\textbf{w}_1(\tau)|\tau,\tilde{\textbf{w}}_1(\tau))\mathrm{d}\textbf{w}_1(\tau)
	\end{aligned}
	\end{equation}
	where $\tilde{\textbf{w}}_1^{\ast}(\tau)$ denotes to a vector of new unseen measured values with the new sample time matrix $\tau_{\ast}$. The term in the integral, the posterior over function, is computed in \eqref{eq:posteriorfun}. The training data set is obtained from the measured value of the output current of the inverter, which is equal to the load's current while there is a single-DG, at $k_1$ to $k_n$. The likelihood of new unseen data $\tilde{w}_1(k_{n+1})$ can be calculated, and it is a Gaussian distribution as well.
	
	\subsection{Tube-Based Robust Model Predictive Control}
	
	By neglecting the disturbances and uncertainties of the real system in \eqref{eq:ModelDisc}, the nominal system will be defined as follows
	
	\begin{equation}
	\tilde{x}(k+1) = A\tilde{x}(k)+B\tilde{u}(k) ; \hspace{5pt}\tilde{x}(0) = \tilde{x}_0,\label{eq:NominalModel}
	\end{equation}
	where $\tilde{x}(k) \in \bar{X}$ and  $\tilde{u}(k) \in \bar{U}$ are the nominal trajectory and the nominal input, respectively. The sets $\bar X$ and $\bar U$ will be defined in a way so that the robust performance is achieved. 
	The MPC controller works in this way: At each MPC shot $t$,  $t \in \{0,\;1,\dots,\;T-N+1\}$, the control law is computed for the period $[t, t+1, \dots,t+N-1]$ by solving an optimization problem. Nevertheless, only the first element of the computed input sequence is applied to the system, and the state of the system evolves for one step accordingly. This process continues for the next shots until the control inputs for the whole control horizon are computed. The control and prediction horizon of MPC are considered to be $T$ and $N$, respectively.
	In our problem, the objective is to ensure that the voltage of the load terminal is tracked in the reference signal $r(t)$. The reference signal is a setpoint while a single-DG unit is available in the islanded microgrid. However, by considering multiple DG units, the reference signal varies as time passes.
	The optimal nominal trajectories of the single-DG can be obtained by solving the following nominal MPC controller. 
	\begin{subequations}
		\begin{align}
		\min_{\tilde{x}_{0|0},\tilde{u}_t} \quad&  \sum_{k=0}^{N-1}{\|\tilde{x}_{k|t}-r(t)\|^2_Q+\|\tilde{u}_{k|t}\|^2_R+\|\tilde{x}_{N|t}-r(t)\|^2_P}\\
		\textrm{s.t.} \quad & \tilde{x}_{k+1|t} = A\tilde{x}_{k|t}+B\tilde{u}_{k|t},\\
		&\tilde{x}_{k|t} \in \bar{X},\tilde{u}_{k|t} \in \bar{U},\label{eq:XUconst}\\
		& (\tilde{x}_{N|t}-r(t)) \in \tilde{X}_f \subseteq \bar{X},\\
        & (\tilde{x}_{0|t} -x_{t})\in S_{K}(\infty),
        \label{eq:Optim_const}
		\end{align}\label{eq:Optim}
	\end{subequations}
    where $\tilde{x}_{k|t}$ and $\tilde{u}_{k|t}$ are known as the predicted state and control input of
	the $t^\text{th}$ MPC shot \cite{limon2010robust}, which give $k$ step ahead prediction of the state and input starting from time-step $t$, and $\tilde u_t=[\tilde{u}_{0|t},\ldots,\tilde{u}_{N-1|t}]$. The matrices $Q$ and $R$ are semi-positive definite and positive definite. The terminal weighting matrix $P$, is a positive definite matrix. $\tilde{X}_f$ represents the terminal set and is designed to ensure the stability condition \cite{rawlings2017model}. The set $S_K(\infty)$ will guarantee to keep the error bounded. The calculation method of $S_K(\infty)$ is discussed at the end of this section. Practically, YALMIP Toolbox is utilized \cite{lofberg2004yalmip} to solve this problem, which uses Quadratic Programming (QP) solver.
	By solving the MPC problem in \eqref{eq:Optim}, the optimal nominal input and the corresponding optimal state trajectory will be found as follows
	\begin{align}
	&	\tilde{u}^\ast = (\tilde{u}^\ast(0),\ldots,\tilde{u}^\ast(T-1))\nonumber\\
	&	=[\tilde{u}^\ast_{0|0},\ldots,\tilde{u}^\ast_{T-N|T-N},\tilde{u}^\ast_{T-N+1|T-N},\ldots,\tilde{u}^\ast_{T-1|T-N}] \nonumber\\
	&	 \tilde{x}^\ast = (\tilde{x}^\ast(0),\ldots,\tilde{x}^\ast(T-1),\tilde{x}^\ast(T)) \nonumber\\
	&=[\tilde{x}^\ast_{0|0},\ldots,\tilde{x}^\ast_{T-N+1|T-N},\ldots,\tilde{x}^\ast_{T-1|T-N},\tilde{x}^\ast_{T|T-N}].
	\end{align}

	Clearly, the real trajectory of the system, $x$, will be different from the nominal trajectory, $\tilde x$, due to the presence of uncertainties and disturbances. The real control input of the system, $u$, is designed in a way that the real trajectory of the system lies close to the nominal trajectory as possible, using the following feedback policy
	
	\begin{equation}
	u(k) = \tilde{u}^\ast(k) + K(x(k)-\tilde{x}^\ast(k)),\label{eq:FeedPol}
	\end{equation}
	where the feedback gain $K \in R^{2\times4}$ can be obtained by pole placement, LQR, and LMI technique to make the matrix $A_K=A+BK$ Schur stable under the controllability of the pair (A,B). In  \eqref{eq:FeedPol}, the gain $K$ compensates deviations from nominal trajectories. By calculating the difference between the real system and nominal system and substituting the feedback policy \eqref{eq:FeedPol} in \eqref{eq:ModelDisc} the error dynamic becomes
	\begin{equation}
	e(k+1) = A_Ke(k)+w(x,u,k),\label{eq:Ak}
	\end{equation} 
	where $e:=x-\tilde{x}^\ast$ and $w(x,u,k) = w_1(k)+w_2(x,u,k)$ whose set is denoted by $W$. As it was mentioned and also will be discussed in the next subsection in more detail, this set is estimated at each iteration using the GP estimator of $w_1(k)$ and the worst-case bound on $w_2(x,u,k)$.
	
	The accumulative set of disturbances, $S_K(k)$, is defined as
	\begin{align}
	S_K(k) := \sum_{j = 0}^{k-1}A_K^j W = W \oplus AW \oplus \dots \oplus A^{k-1}W,\label{eq:W}
	\end{align}
	As shown in \cite{rawlings2017model}, by setting $x(0)=\tilde{x}(0)=x_0$, since $A_K$ is Schur, $S_K(\infty)$ exists and is positive invariant for the error dynamics in \eqref{eq:Ak}.	 The feedback policy \eqref{eq:FeedPol} assures that the state of the real system \eqref{eq:ModelDisc} is compelled to be close to that of the nominal system \eqref{eq:NominalModel}. 	
	
	As $x = \tilde{x}^\ast +e$, and knowing that $K$ is constant, the state of the nominal trajectories will be the center of the tube generated by feedback policy \eqref{eq:FeedPol} is defined as follows
	
	\begin{align}
		x \in Z(k) \coloneqq {\tilde{x}^\ast(k) \oplus S_K(\infty)}
	\end{align}
	
	Finally, to maintain the state and control input in $X$ and $U$, the state and control input set of the nominal system need to be defined as follows
	\begin{equation}
	\begin{aligned}
	\hspace{1mm} \bar{U} \overset{\Delta}{=} U \ominus KS_K(\infty),
	\hspace{1mm} \bar{X} \overset{\Delta}{=} X \ominus S_K(\infty),
	\hspace{1mm} \tilde{X}_f \subset X \ominus S_K(\infty),
	\end{aligned}
	\end{equation}
	
	\begin{proposition}
			Considering $x(k) \in Z(k)$ and $u(k) = \tilde{u}^\ast(k)+K(x(k)-\tilde{x}^\ast(k))$, then $x(k+1) \in Z(k+1)$ for all $w(x,u,k) \in W$ \cite{mayne2005robust}.
	\end{proposition}	
	
	As a result, the solution of \eqref{eq:ModelDisc} using control policy \eqref{eq:FeedPol} as its input lies in the tube $Z(k)$ for every admissible disturbance sequence \cite{rawlings2017model}.
	
	\section{Learning Tube-Based Robust MPC for Islanded Microgrid}\label{section:LRMPC}
	
	In this subsection, we show how the set $W$ and accordingly $S_K(\infty)$ can be estimated using GP and the available bound on $w_2$. This estimation will then be used for tube shaping and formulating the MPC controller of the real system.

	Since the disturbance vector $w_1(k)$ is measured directly, it will be used to collect the training data. We get the sample-time vector of training data as $[k_1,k_2,\cdots,k_n]=[0,\dots,k-1]$ and the prediction sample-time as $k_{n+1}=k$. Therefore, by estimating $\mu_{\ast}(k)$ and $\sigma^2_{\ast}(k)$ using (\ref{eq:posteriorfun}), the distribution of $w_1(k)$ is obtained as GP $\sim \mathcal{N}(\mu_{\ast}(k),\sigma^2_{\ast}(k))$. By assuming ${\left\| {\mu_{\ast} (t + 1) - \mu_{\ast} (t)} \right\|_2} \le \Delta \mu$, the set $\hat{W}$ can be defined as follows which represents an estimation of $W$.
	\begin{equation}
	\begin{aligned}
	\hat{W}(k) = \{&w(x,u,k) \in R^n :\\
	&\|\hat{w}_{1}(k)- 	{\mu_{\ast}(k)} \|_{{\sigma^2_{\ast}(k)}} \le 			\chi^2_{n}(\vartheta)+\Delta \mu,  \\
	&\|w_2(x,u,k)\|_\infty \le L_{2}\},
	\end{aligned}\label{distsethat}
	\end{equation}
	where $\chi^2_n$ and $\vartheta$ denote the chi-distribution with $n$ degree of freedom and the confidence interval. More details are discussed in \cite{bonzanini2021learning}. By substituting $\hat{W}(k)$ into \eqref{eq:W}, the set $\hat{S}_K(\infty)$ can be defined as follows
	
	\begin{align}
	\hat{S}_K{(k)} := \sum_{j = 0}^{k-1}A_K^j \hat{W} = \hat{W} \oplus A\hat{W} \oplus \dots \oplus A^{k-1}\hat{W},\label{eq:hatW}
	\end{align}
	Accordingly, the new state and control input set of the nominal system will be
	
	\begin{equation}
	\begin{aligned}
	\hspace{1mm} \hat{U} \overset{\Delta}{=} U \ominus K\hat{S}_K(\infty),
	\hspace{1mm} \hat{X} \overset{\Delta}{=} X \ominus \hat{S}_K(\infty),
	\hspace{1mm} \hat{X}_f \subset X \ominus \hat{S}_K(\infty),
	\end{aligned}\label{constraints}
	\end{equation}
	In this way, we can replace constraint \eqref{eq:Optim_const} and \eqref{eq:XUconst} in optimization problem \eqref{eq:Optim} with $(\tilde{x}_{N|t}-r(t)) \in \hat{X}_f \subseteq \hat{X}$ and $\tilde{x}_{k|t} \in \hat{X},\tilde{u}_{k|t} \in \hat{U}$, Then  the Learning Tube-Based RMPC optimization problem can be defined as follows:
	\begin{subequations}
		\begin{align}
		\min_{\tilde{x}_{0|0},\tilde{u}_t} \quad&  \sum_{k=0}^{N-1}{\|\tilde{x}_{k|t}-r(t)\|^2_Q+\|\tilde{u}_{k|t}\|^2_R+\|\tilde{x}_{N|t}-r(t)\|^2_P}\\
		&\tilde{x}_{k+1|t} = A\tilde{x}_{k|t}+B\tilde{u}_{k|t},\label{eq:nom_lbmpc}\\
		&\tilde{x}_{k|t} \in \hat{X},\tilde{u}_{k|t} \in \hat{U},\\
		& (\tilde{x}_{N|t}-r(t)) \in \hat{X}_f \subseteq \hat{X}, \\
        & (\tilde{x}_{0|t} -x_{t})\in \hat{S}_{K}(\infty),
  \label{eq:Finalstate}
		\end{align}\label{eq:OptimFinal}
	\end{subequations}
	where $\hat{X}_f$ is the invariant terminal set which is computed using the disturbance set introduced in \eqref{distsethat} and the method presented in \cite{fagiano2012model}. 	
   The procedure of designing Learning RMPC is illustrated in Algorithm \ref{Algorithm}.
    
 \begin{algorithm}[h]
		\nl \KwIn{initial DataSet and GP, $t=t_0$ \bf Measure \(x_t\) \;
		\nl \bf Update GP's DataSet and compute \({\mu}_{*}\) and \({\sigma}_{*} \) \;
		\nl \bf Update tube and constraint by using Equ. (\ref{distsethat}), (\ref{eq:hatW}), (\ref{constraints})  \;
		\nl \bf Set \({x}_{0|t} = x_t\) \;
		\nl \bf Solve Problem (\ref{eq:OptimFinal})   \;
		\nl \bf Set \(\tilde{u}_{t} = \tilde{u}_{0|t} \)   \;
		\nl \bf Implement \(u_{t} = \tilde{u}_{t} + K (x_{t}-\tilde{x}_{0|t})\) \;
		\nl \bf Set \(t = t+1\) and go to step 1;}
		\caption{{\bf Learning Tube-Based RMPC } \label{Algorithm}}
\end{algorithm}

	\section{Analytical Results}\label{section:AS}

	\subsection{Recursive Feasibility}
	
	
        An MPC is called recursively feasible if it always keeps the states in a region from where the online optimization problem has a feasible solution \cite{8920443}. In other words, if the problem (\ref{eq:OptimFinal})
        has a feasible solution for the initial condition $x_0$ and it remains feasible for any subsequent states $x_i$ of the controlled system (3), then the MPC problem is recursively feasible \cite{book}.
        
	\begin{lemma}
		Let's define ${u}_{r} = B^{\dagger}(r-A {r})$. Then by having  $\tilde{u}_{N|t} = K(\tilde{x}_{N|t}-r)+u_{r}$, an invariant terminal set $ \hat{X}_f\ $ exists such that if  $ \ \tilde{x}_{N|t}-r \in \hat{X}_f\ $ then $A\tilde{x}_{N|t}+B\tilde{u}_{N|t}-r\in \hat{X}_f$.
	\end{lemma}

\textit{Proof:} proof is provided in \cite{fagiano2012model}.

	Now we can express the Recursive Feasibility theorem. 
	\\
	
    \begin{theorem}. Assume that $\hat{X}_f$ is a terminal invariant set given in Lemma 1. If the MPC problem has a feasible solution for the initial condition $x_0$, then the solution of MPC is feasible for all the times.
    \end{theorem}
	
    \textit{Proof}:
    The proof is based on mathematical induction. 
    For this purpose, it is assumed that the optimization problem \eqref{eq:OptimFinal} at time $t$ has a solution like ([$\tilde{x}^\ast_{0|t}, \tilde{x}^\ast_{1|t},\cdots,\tilde{x}^\ast_{N|t}$],[$\tilde{u}^\ast_{0|t}, \tilde{u}^\ast_{1|t},\cdots,\tilde{u}^\ast_{N-1|t}$]).
	
	Since the tube $Z(k)$ is an invariant set for the error dynamics (\ref{eq:Ak}), therefore,
	$([ \tilde{x}^\ast_{1|t}, \tilde{x}^\ast_{2|t}, \tilde{x}^\ast_{3|t},\cdots,\tilde{x}^\ast_{N|t}$] , [$\tilde{u}^\ast_{1|t}, \tilde{u}^\ast_{2|t},\cdots,\tilde{u}^\ast_{N-1|t}])$
	is feasible solution for MPC problem at $t+1$ from $1$ to $N-1$ prediction horizon.
	Also, according to the constraint $(\tilde{x}_{N|t}-r) \in \hat{X}_f$ and Lemma 1, we can say that there is an input $\tilde{u}$ and final state $\tilde{x}_{N|t+1}$ that satisfies the equation \eqref{eq:Finalstate}. Hence, a feasible solution for optimization \eqref{eq:OptimFinal} at time $t+1$ can be suggested as follows 
	\begin{equation}
	\begin{aligned}
	(\boldsymbol{x}_{t+1},\boldsymbol{u}_{t+1})=&([\tilde{x}^\ast_{1|t}, \tilde{x}^\ast_{2|t},\cdots,\tilde{x}^\ast_{N|t},\\ &A\tilde{x}_{N|t}+B({K}(\tilde{x}_{N|t}-r)+u_{r})] \\ 
	&,[\tilde{u}^\ast_{0|t}, \tilde{u}^\ast_{1|t},\cdots,\tilde{u}^\ast_{N-1|t}
	, {K}(\tilde{x}_{N|t}-r)+u_{r}])
	\label{eq:nextsol}
	\end{aligned}
	\end{equation}
	Therefore, the proof is completed by mathematical induction.
	\subsection{Stability}
	It is also important to verify the stability of the system under the proposed Learning RMPC algorithm. The following theorem expresses the stability of the system.
	
	\begin{theorem}
	    Using control policy \eqref{eq:FeedPol} with the nominal input derived from the optimization problem \eqref{eq:Optim} subject to the system \eqref{eq:ModelDisc}, the following properties are satisfied:
	
	\begin{enumerate}
		\item $x(k) \in X$, $u(k) \in U$ for all $k \geq 0$.
		\item The closed-loop system is Input-to-State-Stable (ISS).
	\end{enumerate}
	\end{theorem} 
	
	\textit{Proof.} In the following, by using the Lyapunov method, the stability of the system will be expressed.
	
 $J(t)$ is defined as the cost function used in the learning tube-Based RMPC controller as follows
	\begin{equation}
	\begin{aligned}
	J(t)\ = \sum_{k=0}^{N-1}{\|\tilde{x}_{k|t}-r\|^2_Q+\|\tilde{u}_{k|t}\|^2_R+\|\tilde{x}_{N|t}-r\|^2_P}\ 
	\end{aligned}\label{costf}
	\end{equation}
then, the Lyapunov function can be defined as the optimal cost function obtained from \eqref{eq:Finalstate} as follows
	\begin{equation}
	\begin{aligned}
	v_{t} = J_\ast(t)
	\label{eq:lyapunovdef}
	\end{aligned}
	\end{equation}
	$v_{t}$ is clearly positive. According to the Lyapunov theorem, if the Lyapunov function can be proved to be decreasing over the time, then the system is stable. 
	\begin{equation}
	\begin{aligned}
	v_{t+1}-v_{t} = J_\ast(t+1)-J_\ast(t)  
	\label{eq:stability1}
	\end{aligned}
	\end{equation}
	where $J_\ast(t+1)$ is the optimal cost at $t+1$. Let $\hat{J}(t+1)$ be the cost function obtained by substituting $u_{t+1}$ in \eqref{costf}, then:
	\begin{equation}
	\begin{aligned}
	v_{t+1}-v_{t} &\leq \hat{J}(t+1)-J_\ast(t) \\
	&\leq   \|\tilde{x}_{N|t}-r\|^2_Q+\|\tilde{u}_{N|t}\|^2_R+\|\tilde{x}_{N|t+1}-r\|^2_P\\
	&-(\|\tilde{x}_{0|t}-r\|^2_Q+\|\tilde{u}_{0|t}\|^2_R+\|\tilde{x}_{N|t}-r\|^2_P)
	\label{eq:stability2}
	\end{aligned}
	\end{equation}
	
	The $\|\tilde{x}_{N|t+1}-r\|^2_P$ can be simplified as follows
	\begin{equation}
	\begin{aligned}
	&\tilde{x}_{N|t+1}-r \\
	&=A\tilde{x}_{N|t}+B({K}(\tilde{x}_{N|t}-r)+u_{r})-r\\
	&=(A+B{K})(\tilde{x}_{N|t}-r_{k})
	+Ar+Bu_{r}-r\\
	&=(A+B{K})(\tilde{x}_{N|t}-r)\\
	&\implies \|\tilde{x}_{N|t+1}-r\|^2_P
	\leq \|  (A+B{K})(\tilde{x}_{N|t}-r) \|^2_P 
	\label{eq:stability3}		
	\end{aligned}
	\end{equation}
	by considering \eqref{eq:stability2} and \eqref{eq:stability3} :
	\begin{equation}
	\begin{aligned}
	v_{t+1}-v_{t} \leq 
	&\|\tilde{x}_{N|t}-r\|^2_Q+\| {K}(\tilde{x}_{N|t}-r)  \|^2_R\\ 
	+& {\| u_{r} \|^2_R}+\|  A_{ {K}}(\tilde{x}_{N|t}-r) \|^2_P \\
	-&(\|\tilde{x}_{0|t}-r\|^2_Q+\|\tilde{u}_{0|t}\|^2_R+\|\tilde{x}_{N|t}-r\|^2_P)\\
	\leq &\|\tilde{x}_{N|t}-r\|^2_{Q+ {K}^{T}R {K}+(A_{ {K}})^TP(A_{ {K}})-P}\\
	+ &{\| u_{r} \|^2_R}-(\|\tilde{x}_{0|t}-r\|^2_Q+\|\tilde{u}_{0|t}\|^2_R)  
	\label{eq:stability5}
	\end{aligned}
	\end{equation}
	${K}$ should be chosen in such a way that it applies to the discrete Lyapunov equation $ P=Q+{K}^{T}R{K}+(A_{{K}})^TP(A_{{K}}) $, then:
	\begin{equation}
	\begin{aligned}
	v_{t+1}-v_{t} \leq {\| u_{r} \|^2_R}-(\|\tilde{x}_{0|t}-r\|^2_Q+\|\tilde{u}_{0|t}\|^2_R)
	\label{eq:stability6}
	\end{aligned}
	\end{equation}
	\eqref{eq:stability6} is negative when  $(\|\tilde{x}_{0|t}-r\|^2_Q) \leq {\| u_{r} \|^2_R}$ . Therefore, the system (\ref{eq:ModelDisc}) with the proposed control design within the region of attraction $(\|\tilde{x}_{0|t}-r\|^2_Q) \leq {\| u_{r} \|^2_R}$ is stable.

	\section{Simulation Results}\label{section:SR}

To evaluate the performance of the proposed Learning Tube-Based RMPC, first, a single-DG microgrid shown in Fig. \ref{fig:DGunit} is simulated to show the efficiency of voltage control. Then, two DGs are connected in parallel and power sharing has been done using droop control. The microgrid parameters are listed in Table \ref{tab1}.

\begin{table}[H]
	\begin{center}
		\caption{Parameters of DG}
		\label{tab1}
		\begin{tabular}{ c  c | c }
			\hline \hline
			Parameter    & Value & Remark \\
			\hline
			$R_f$      & 1.5 m$\Omega$   &     \\
			$L_f$       & 100 mH     &     \\			
			$C_f$       & 100 $\mu$F     &      \\
			$f_0$       & 60 Hz    & Converter and    \\
			$S_{base}$       & 3 MVA    &  LC Filter    \\
			$V_{base}$       & 600 V    &     \\
			$V_{dc}$       & 2000 V    &     \\
			Transformer ratio ($Y$/$\Delta$)      & 600/13800 V    &     \\
			\hline
			$T_s$      & 250 $\mu$s   & MPC     \\
			\hline
                 $Z_{Line1}$, $Z_{Line2}$ & 0.35+j1.16 $\Omega$ & \\
                $n_1$,$n_2$ & 0.5, 0.87 V/MVAr  & Droop Parameters \\
                $m_1$,$m_2$ & 0.6, 0.9 Hz/MW & \\
                 \hline
		\end{tabular}
	\end{center}
\end{table}

Tube-Based RMPC and Learning Tube-Based RMPC are used for the DG in the same operational condition. A comparison between methods has been made to illustrate the priority of the proposed method. During all the simulations, The upper bound of parametric uncertainties is considered in parameters while it is assumed that $\Delta R_f$ = 0.1$R_f$, $\Delta C_f$ = 0.1$C_f$ and $\Delta L_f$ = 0.2$L_f$. THD, tracking with low steady-state error and proper dynamic response as evaluation features, has been used in this comparison. Also, parasitic elements of switches and filter, besides the sampling and computational delays of 202$\mu$s, have been considered. 
THD permissible range is defined by the IEEE standard \cite{6826459} (i.e., 5\%). Simulations have been done by the use of Matlab, Python, YALMIP, and SimPowerSystems toolbox.

The proposed method shows how much variation exists in the worst case by assuming a bound for disturbances and parametric uncertainties shaping the tubes. The tubes are calculated online during the operation, which will be advantageous for precise analysis. Shaping the control policy mentioned in \eqref{eq:FeedPol} will be a crucial step in Tube-Based RMPC. It can be proven that, under the controllability of (A,B) and observability of (A,$\sqrt{Q}$), the gain $K$ obtained from above LQR ensures the {optimality} and {stability} of the system \cite{Bert05}. Compared to other gains, the chosen $K$ will minimize the control effort and give a better transient response. Using the LQR technique, matrix $K$ is obtained as follows

\begin{align}
K& =
\begin{bmatrix}
-2.79 \times 10^{-4} & -1.14 \times 10^{-4} & 0.0369 & 0.018 \\
-0.0028 & 0.0057 & 0.018 & 0.1141 \\
\end{bmatrix},
\end{align} 

If $K$ does not determine correctly, at least two problems will happen: First, the controller will not be able to track the voltage precisely. Second, it is probable that the system will become unstable. noisy observations are considered, which can represent the measurement noise with $\delta^2_{\varepsilon}$ = 0.01 variance. Moreover, a 95\% confidence interval will be shaped for the noisy data, which will be used as a bound to shape the tubes. It is worth mentioning that the proposed method can be implemented practically using MBE.300.E500 PMSM, commercially available by Technosoft SA. Considering the prediction horizon of 1.25ms, and the control horizon of 85ms, it takes 0.027455s to run Learning Robust Model Predictive Control for the proposed model on average.




\subsection{Harmonic Loads}

\begin{figure}[t]
	\begin{center}
		\includegraphics[width=9.5cm]{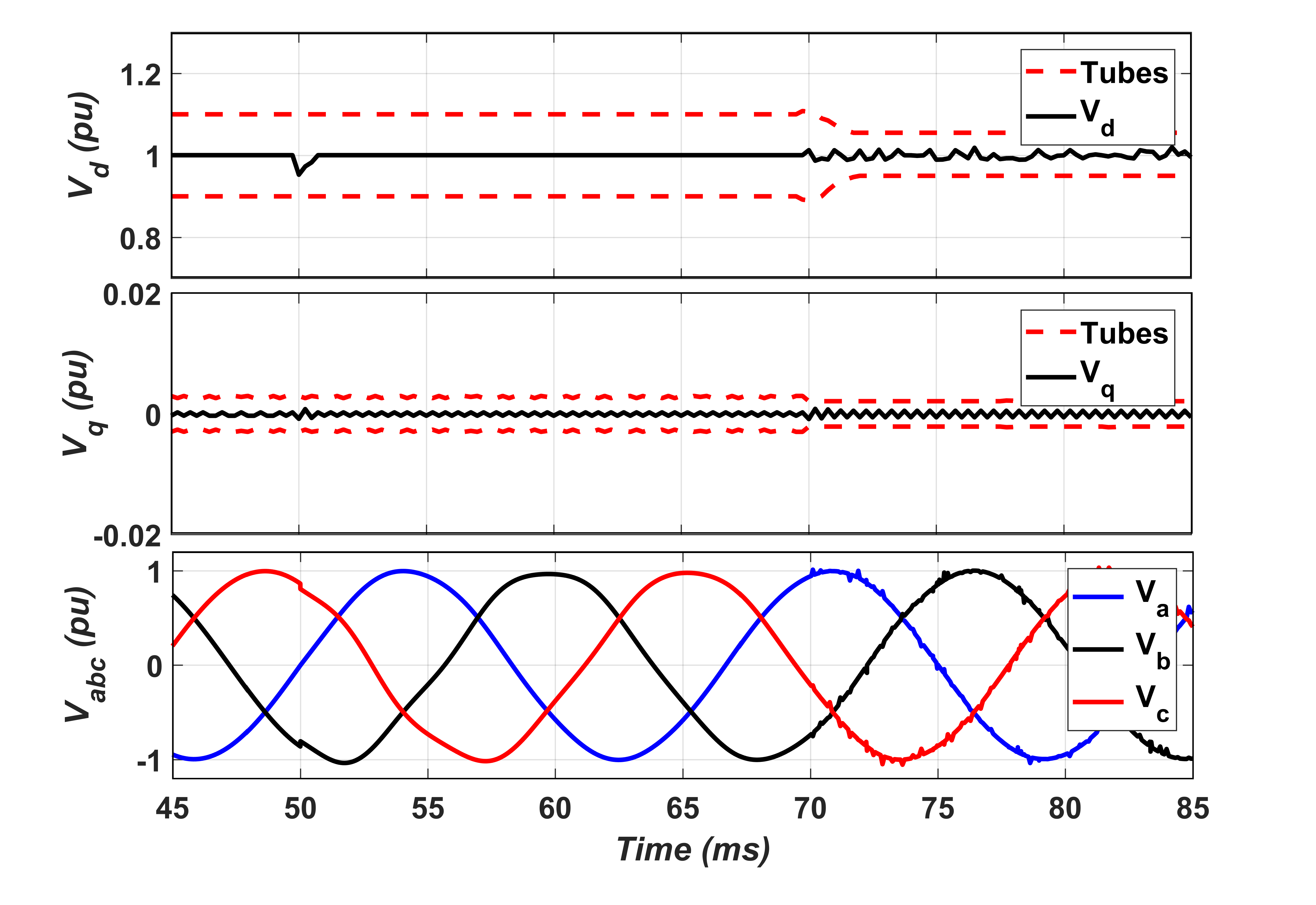}
		\caption{Voltage regulation using Learning Tube-Based RMPC: voltages in d-q frame and abc-frame while the constant impedance and harmonic loads are connected at $t$ = 50 ms and $t$ = 70 ms.}
		\label{fig:LRMPCcompact}
	\end{center}
\end{figure}
In this section, the performance of the proposed method is evaluated. At $t$ = 50ms, a 340kVA constant impedance load with PF = 0.9 is connected to the islanded microgrid, and a nonlinear load consisting of 5th and 7th harmonics is connected at $t$ = 70ms. The THD of the nonlinear load's current is approximately 38\%.

As shown in Fig. \ref{fig:LRMPCcompact}, voltage regulation has been done efficiently using Learning Tube-Based RMPC. The THD of the voltage is 2.01\%, which is desirable. The tubes get tighter when the learning method is used at $t$ = 70ms. 

\begin{figure}[H]
	\begin{center}
		\includegraphics[width=8.5cm]{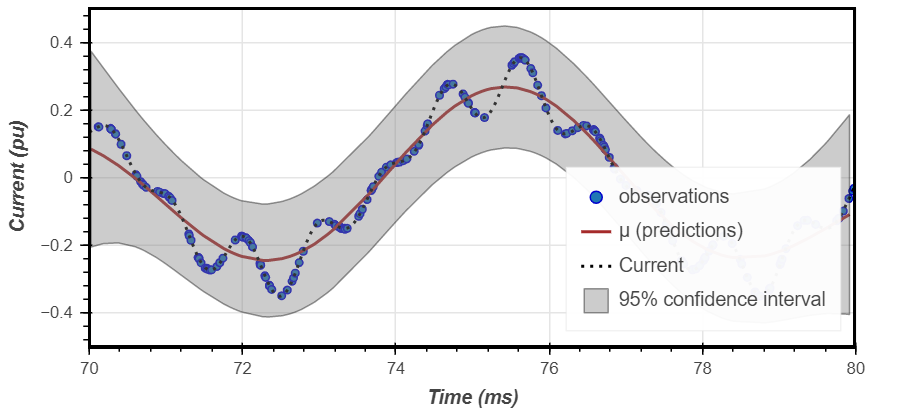}
		\caption{Gaussian Process Regression is used to predict the harmonic current and shape a 95\% confidence interval with noisy observations.}
		\label{fig:Noisy157}
	\end{center}
\end{figure}

In Fig. \ref{fig:Noisy157}, estimating the current of the load has been shown with the 95\% confidence interval. The performance is enhanced because $\mu_{\ast}$, as an estimation, and $\sigma_\ast^2$ is used to give the bounds besides the parametric uncertainties upper bound, which is calculated and implemented in tubes.

\begin{table}[H]
	\renewcommand{\arraystretch}{1.3}
	\caption{Comparison Among Methods for Different Harmonic Loads}
	\label{table2}
	\centering
	\begin{tabular}{|c|c|c|c|c|}
		\cline{3-5}
		\multicolumn {2}{c|}{} & \multicolumn {3}{c|}{\textbf{Voltage THDs} $\%$}\\ \hline
		\textbf{Harmonics}     & \textbf{Current THDs} $\%$ & \textbf{LRMPC} & \textbf{RMPC} & \textbf{MPC} \\ \hline
		\multirow{3}{*}{5$^{th}$, 7$^{th}$, 11$^{th}$}  & 30.83 & 1.97      & 2.02 & 3.58     \\ \cline{2-5}
		& 41.69 & 2.03      & 2.07   & 3.67   \\ \cline{2-5}
		& 51.39 & 2.07      & 2.15 & 3.91\\ \hline
		\multirow{3}{*}{5$^{th}$, 11$^{th}$, 13$^{th}$} &31.49  & 2.02      & 2.11  & 3.62    \\ \cline{2-5}
		& 44.62 & 2.12     & 2.24  & 3.76    \\ \cline{2-5}
		& 52.71 & 2.21     & 2.36  & 4.02 \\ \hline
	\end{tabular}
\end{table}

In Table \ref{table2}, various harmonic loads have been connected to the islanded microgrid, and regulating the voltage has been done. Voltage THDs have been calculated and reported. As the table illustrates, voltage THDs are desirable for the proposed method. It is worth mentioning that the conventional controller, PI, have been implemented to handle the above scenario while the harmonic load with $5^{th}$, $7^{th}$, and $11^{th}$ and the current THD of 38\% is connected. The results have been shown in the following table

\begin{table}[H]
	\renewcommand{\arraystretch}{1.4}
	\caption{Comparison Among Methods Based on Voltage THDs - Harmonic Loads}
	\label{tab:table3}
	\centering
	\begin{tabular}{|c|c|c|c|}
		\hline
		Learning RMPC      & Tube-Based RMPC & MPC & Conventional PI \\ \hline
		2.01\% & 2.08\% & 3.62\% & 3.74\\
		\hline
	\end{tabular}
\end{table}

Moreover, the relationship between the learning results and the control input have been illustrated in the following figure

\begin{figure}[H]
	\begin{center}
		\includegraphics[width=9.5cm]{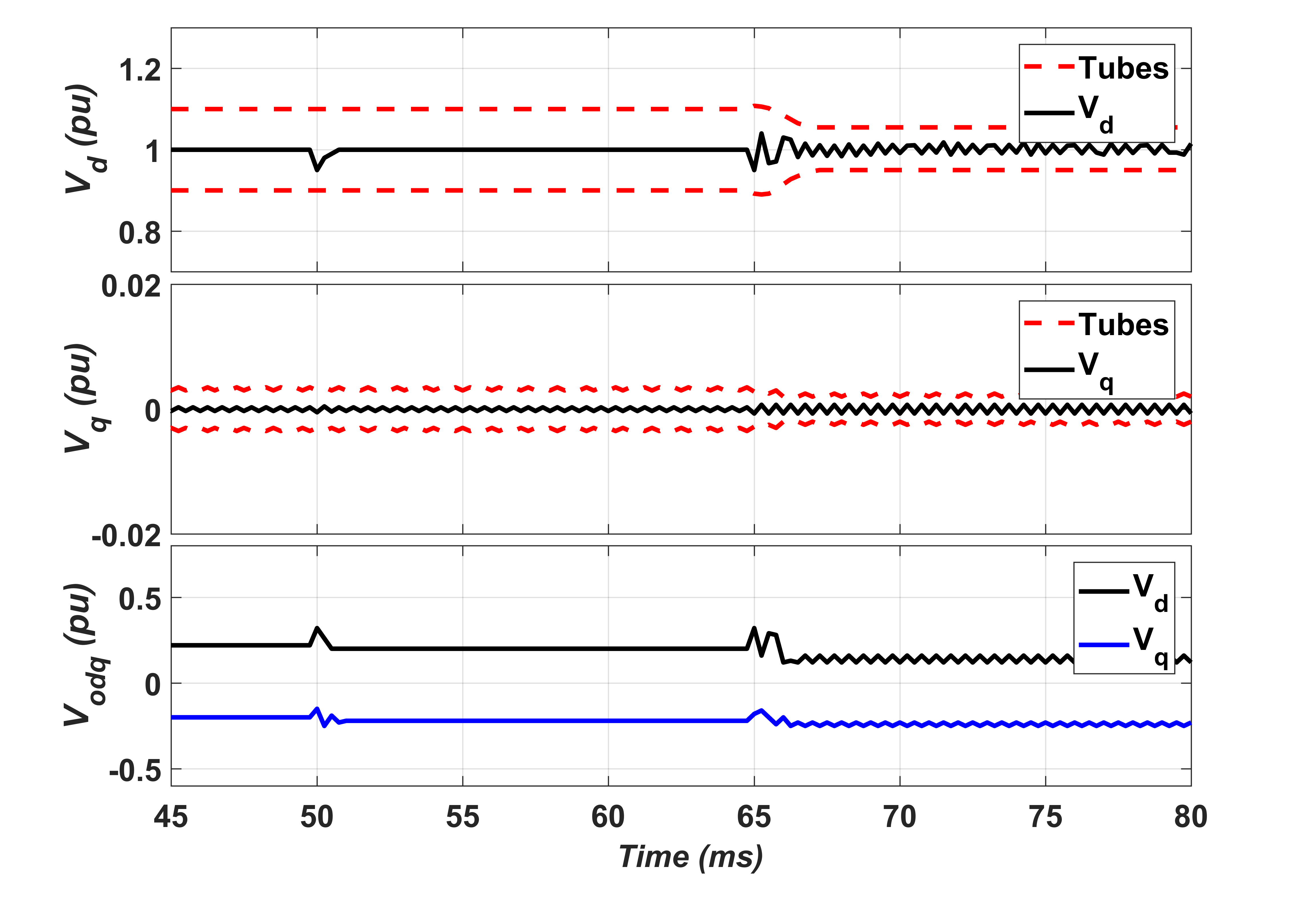}
		\caption{Voltage regulation using Learning Tube-Based RMPC: learning results and control input in d-q frame and abc-frame while the constant impedance and harmonic loads are connected at $t$ = 50 ms and $t$ = 65 ms.}
		\label{fig:SimControl}
	\end{center}
\end{figure}

In Fig. \ref{fig:SimControl}, we have considered a constant impedance load besides the harmonic loads with $5^{th}$, $11^{th}$, and $13^{th}$ and the current THD of 44.62\%.


\subsection{Constant Power Loads}

In this section, a 340kVA constant impedance load with PF = 0.9 is connected to the islanded microgrid at $t$ = 50ms, and a 500kVA constant power load with PF = 1 parallel with a harmonic load consist of $5^{th}$ and $7^{th}$ harmonics is connected to the output terminal at $t$ = 70ms. Learning RMPC is used to handle the voltage regulation.

\begin{figure}[H]
	\begin{center}
		\includegraphics[width=9.5cm]{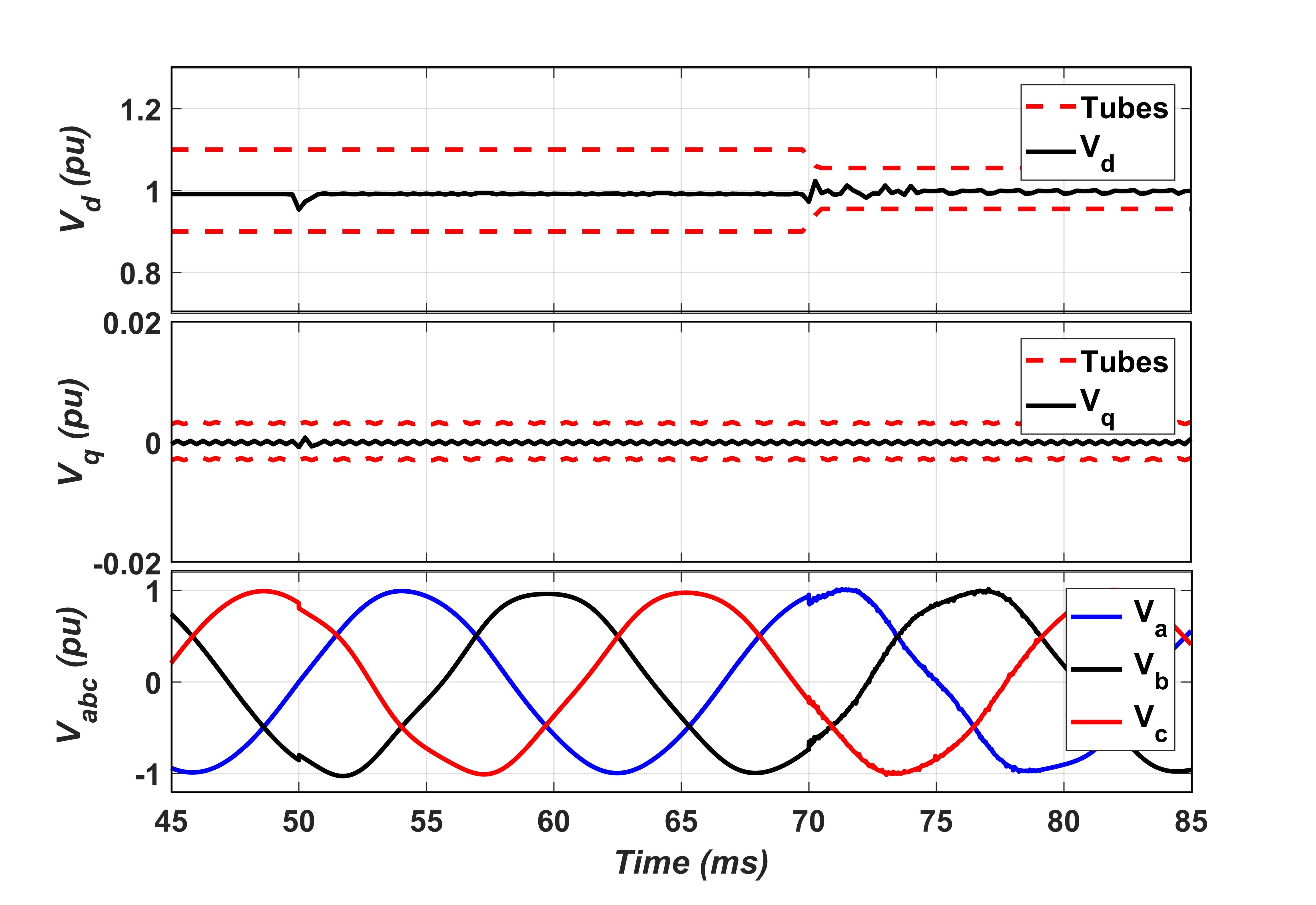}
		\caption{Voltage regulation using Learning Tube-Based RMPC: voltages in d-q frame and abc-frame while the constant impedance and the constant power load parallel with harmonic load are connected at $t$ = 50 ms and $t$ = 70 ms.}
		\label{fig:LRMPCCP}
	\end{center}
\end{figure}

In Fig \ref{fig:LRMPCCP}, Learning RMPC will track the voltage with desirable voltage THD of 2.39\% which is proper. A comparison has been made in the following table among different methods as follows

\begin{table}[H]
	\renewcommand{\arraystretch}{1.3}
	\caption{Comparison Among Methods Based on Voltage THDs - Constant Power Load}
	\label{tab:table4}
	\centering
	\begin{tabular}{|c|c|c|}
		\hline
		Learning RMPC      & Tube-Based RMPC & Conventional PI \\ \hline
		2.39\% & 2.48\% & 3.95\% \\
		\hline
	\end{tabular}
\end{table}
\begin{figure}[H]
	\begin{center}
		\includegraphics[width=8.5cm]{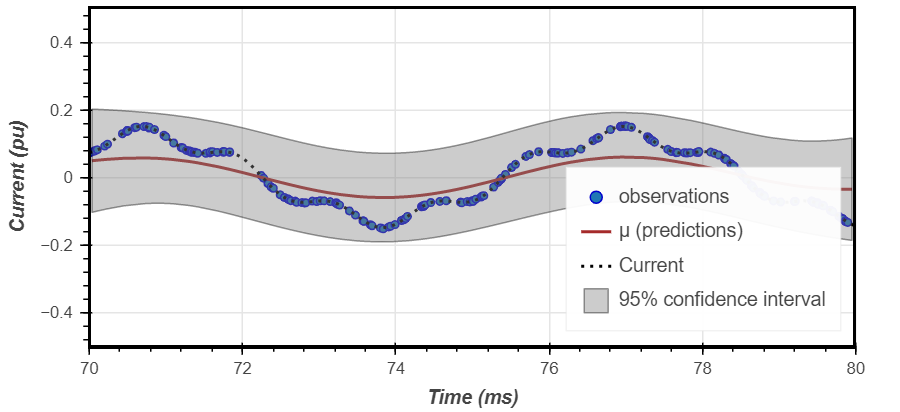}
		\caption{Gaussian Process Regression is used to predict the current of the loads and shape a 95\% confidence interval with noisy observations.}
		\label{fig:GPCP}
	\end{center}
\end{figure}

In Fig. \ref{fig:GPCP}, GP is used to predict the current of the loads. The variance has been used to shape the tube with less conservativeness.

\subsection{Droop Control for Power Sharing}

Learning RMPC is proposed at the primary level of control for a single-DG unit. The proposed method can be also used in a microgrid with multiple DG units. two DGs are considered for adding the power sharing layer while they are working in parallel using droop control method.
    
 
    The instantaneous voltage and the current of the output terminal will be used to calculate the active power
    and reactive power, respectively. Then, the powers will be applied to a low pass filter with a cutoff frequency of 10Hz to remove the ripples. $P_{DG}$ and $Q_{DG}$ as the average powers will be the input of droop controller to produce the reference signals as follows
	
	\begin{equation}
	\begin{aligned}
	&f_i = f_{n_i}-m_{i}P_{DG_i}, \\
	&v^*_{d_i} = V_{n_i}-n_{i}Q_{DG_i}, \\
	&v^*_{q_i} = 0,
	\label{eq:droop}
	\end{aligned}
	\end{equation}
	where $m_{P_i}$ and $n_{Q_i}$ are the frequency and voltage droop gains, respectively. $V_{n_i}$ and $f_{n_i}$ are the desired frequency and voltage for i-th DG unit, respectively.
 
Droop control is a vital aspect of parallel inverter operation, while phase-lock loops (PLL) are commonly used for synchronization. Notably, droop control possesses an inherent synchronization mechanism, obviating the need for a dedicated synchronization unit. This allows droop control to effectively achieve the same function as a PLL, as detailed in the study by Zhong and Zhang (2020) \cite{zhong2020synchronization}. 

\begin{figure}[H]
	\begin{center}
		\includegraphics[scale=0.125]{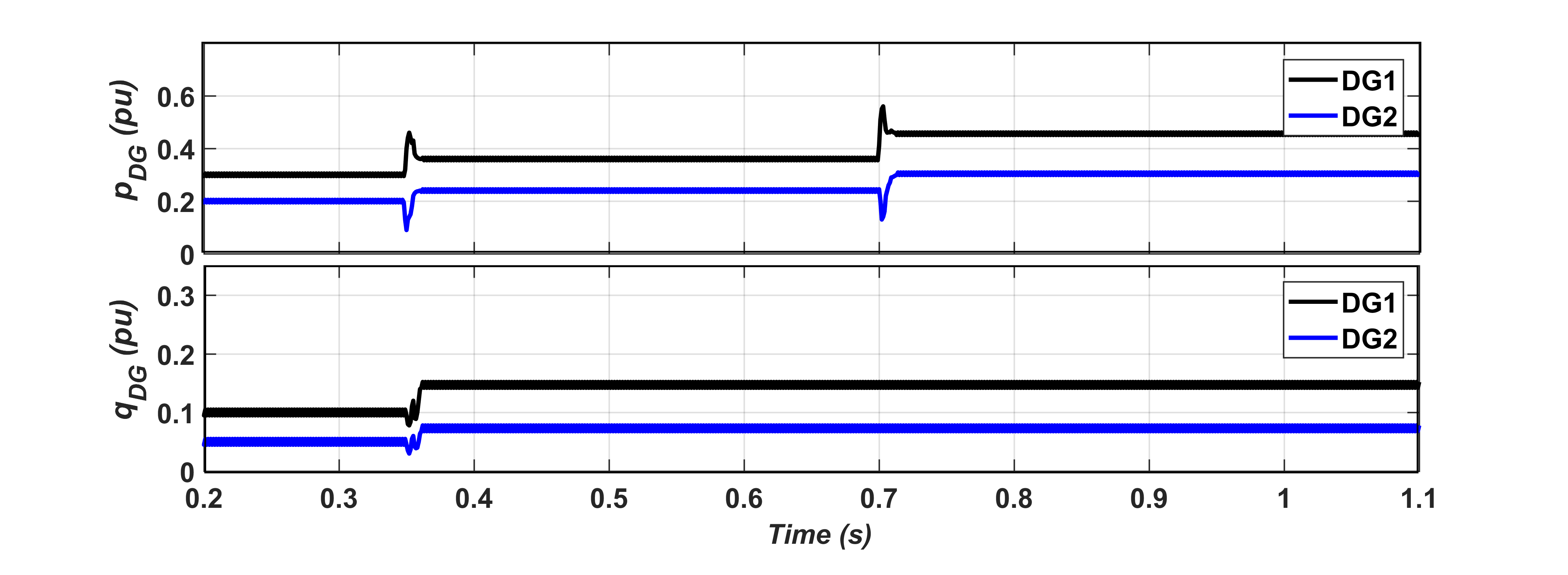}
		\caption{Instantaneous power and reactive power of two DG units while a constant impedance and a constant power load are connected.}
		\label{fig:Droop}
	\end{center}
\end{figure}

Fig. \ref{fig:Droop} shows the instantaneous active and reactive powers while a 340kVA constant impedance with PF=0.9 and a 500kVA constant load with PF=1 are connected at $t$ = 350ms and $t$=700ms, respectively. As observed, the LPF with $f_L$ = 10Hz removes the ripples, and the average of powers is shared among DG units among them according to their droop charactristics.

\begin{figure}[H]
	\begin{center}
		\includegraphics[scale=0.125]{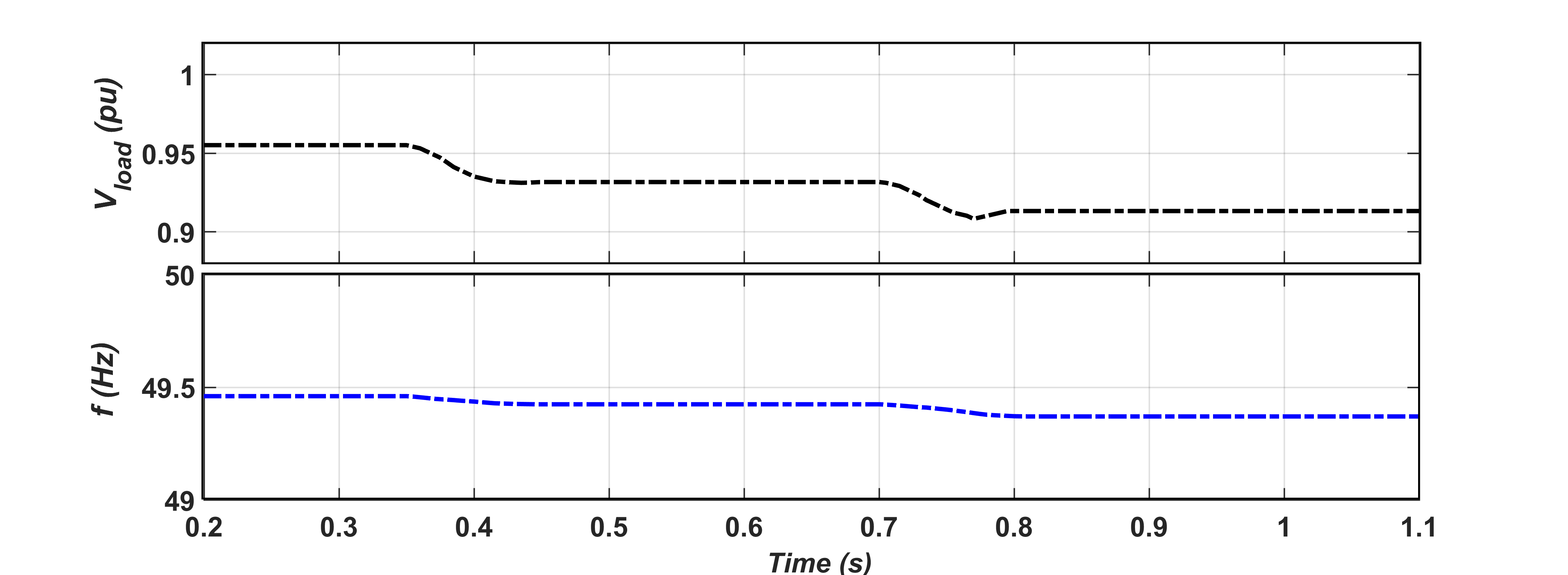}
		\caption{Voltage and frequency of the load terminal.}
		\label{fig:VF}
	\end{center}
\end{figure}
Fig. \ref{fig:VF} shows the voltage and frequency of the load terminal. The references from droop controller will be used in Learning RMPC to force to track them. As the dynamics are slow, the performance of LRMPC is not affected significantly.

	\section{Conclusions}\label{section:Conc}

    This paper proposed a novel Learning Tube-Based RMPC to track the voltage of an AC inverter-based islanded microgrid at the primary level of control. This technique could handle nonlinear loads as disturbances—besides the upper bound of parametric uncertainties. As a powerful learning method, GPR assisted in designing the Tube-Based RMPC with less conservativeness by estimating the load and shaping the tubes online using available data.
    The advantages and disadvantages of the proposed method are mentioned as follows:
    (i): By using Tube-Based RMPC, the performance of voltage regulation of an islanded microgrid is enhanced in the presence of parametric uncertainties and load disturbances to meet PQ standards. (ii) The performance of Tube-based RMPC is enhanced by incorporating a Gaussian Process method to estimate the mean value and the bound of uncertainties based on real data, which leads to less conservativeness compared to the worst-case design. (iii): Harmonic loads as a disturbance are handled using Learning Tube-Based MPC. (iv): The computational cost of Learning Tube-Based MPC is increased compared to benchmarking methods like PI, MPC, and also Tube-Based MPCs.
    
    Moreover, we showed that our method can be used in an islanded microgrid with multiple DG units, while droop control was used for power sharing among DGs. Besides, the recursive feasibility and stability analysis of Learning Tube-Based RMPC has been verified analytically. Finally, MPC, Tube-Based RMPC, and Learning Tube-Based RMPC were compared with each other in terms of THD, performance, and steady-state error in different conditions, which verified the effectiveness of the proposed Learning Tube-Based RMPC.
	
	\bibliography{ref}
	
\end{document}